\documentclass{article}

\PassOptionsToPackage{numbers,sort&compress}{natbib}

 \usepackage[preprint]{neurips_2026}

\usepackage[utf8]{inputenc} % allow utf-8 input
\usepackage[T1]{fontenc}    % use 8-bit T1 fonts
\usepackage[hidelinks]{hyperref}       % hyperlinks
\usepackage{url}            % simple URL typesetting
\usepackage{booktabs}       % professional-quality tables
\usepackage{amsmath}        % advanced math and \text in math mode
\usepackage{amsfonts}       % blackboard math symbols
\usepackage{nicefrac}       % compact symbols for 1/2, etc.
\usepackage{microtype}      % microtypography
\usepackage{xcolor}         % colors
\usepackage{xspace}         % smart spaces after macros
\usepackage{graphicx}

\usepackage{algorithm}
\usepackage{algpseudocode}

\newcommand{\MF}{MF\xspace}
\newcommand{\MB}{MB\xspace}
\newcommand{\phaseoneep}{200\xspace}
\newcommand{\fig}{Fig.\xspace}
\newcommand{\algr}{Algorithm\xspace}
\newcommand{\algrs}{Algorithms\xspace}
\newcommand{\tabref}{Table\xspace}

\newcommand{\edge}[3]{e_{#1}(#2, #3)}
\newcommand{\memcap}{m}
\newcommand{\treedepth}{d}
\newcommand{\avg}[1]{\left< #1 \right>}

\title{Cortex and subcortex play distinct roles over learning when cortical memory is limited}
\author{
	Matthew Farrell$^{1}$\thanks{Correspondence to: \texttt{matthew.farrell@riken.jp}}
	 $\,$ and Taro Toyoizumi$^{1,2}$\\
	\\
  $^1$Laboratory for Neural Computation and Adaptation\\
  RIKEN Center for Brain Science, Wako, Japan \\
	\\
$^2$Department of Mathematical Informatics,\\
Graduate School of Information Science and Technology,\\
The University of Tokyo, Tokyo, Japan.
}
\begin{document}
\maketitle

\begin{abstract}
  It has been proposed that the brain integrates flexible, computationally expensive cortical processing with simpler, lower-cost subcortical mechanisms to achieve resource-efficient performance greater than that of either system alone. Despite the allure of this perspective, satisfying theoretical frameworks that explore this hypothesis are still limited. We extend existing frameworks in which a model-based module and model-free module learn in tandem by explicitly constraining the memory resources of the model-based module, and investigate the impact of this constraint in a simple decision-making setting. Memory constraints naturally give rise to \emph{strategies} for allocating memory resources. We evaluate the performance of different strategies in different situations and demonstrate that when the rewarded states change often, it can be advantageous for the model-based module to focus its memory resources not on exploiting the current reward, but on capturing general structure of the environment. This work provides a theoretical foundation for a functional dissociation between cortical and subcortical systems during learning: the cortex supports general structure learning, while subcortical circuits specialize in reward-based learning. We further detail how these hypotheses can be tested on experimental data.
\end{abstract}

\section{Introduction}
Central to natural and artificial intelligence (AI) is a trade-off between the power of
a computational module and the amount of resources needed to support it. This trade-off
is well-demonstrated by animals: higher-order thinking supported by (prefrontal) cortex
demands strict attention and memory \citep{otto_curse_2013}, while skills that are
well-learned such that they become nearly automatic require far fewer mental resources
\citep{taylor_role_2012, anderson_acquisition_1982, haith_multiple_2018} and may not
even require cortex to perform \citep{kawai_motor_2015}. The brain thus presents an
intriguing example of a computational machine where ``System 1'' (fast, subconscious,
automatic) and ``System 2'' (slow, deliberate, flexible) modules are thought to work
together during task learning to achieve more than each can do alone. A better
understanding of how this is achieved would yield a better understanding of the brain
as a whole. Since our work touches on how humans make decisions, this research could
eventually be used to better treat psychological conditions such as addiction. Our work
may also contribute to the development of more efficient AI which can reduce power and
other resource consumption. If misused, a better understanding of decision-making can
be used to manipulate people in a way that is harmful.

Many existing works spanning (computational) neuroscience and AI posit such a
two-module system. One popular approach layers model-based (MB) mechanisms on top of a
model-free (MF) learner, which speeds up learning compared to using a \MF learner
alone. A common example is \MB trajectory planning used to guide or augment data
collection for the \MF learner to use. Among such methods, some compute long, carefully
optimized trajectories that led to high reward regions, either in the case where
environment dynamics are known to the agent \citep{levine_guided_2013} or learned
\citep{levine_learning_2014}. Others instead perform shorter, more local planning steps
from the current situation, an approach that has scaled successfully to superhuman
performance on difficult tasks
\citep{silver_mastering_2016,silver_general_2018,schrittwieser_mastering_2020,hafner_dream_2020}.

This idea of trajectory rollouts is also highly relevant in models of the brain. A
specific example of this is hippocampal replay, the reactivation of neurons associated
with past experiences. This observation has led to a plethora of models showing how
such reactivation can be used to improve learning. For example, in
\citep{mattar_prioritized_2018, jensen_recurrent_2024} an agent exploring an
environment is given the opportunity to simulate experiences via replay instead of
taking a real action in the world. Faster learning is achieved by choosing replay
trajectories wisely and balancing replay with real-world interaction.
Neuroscience-inspired replay is also influential in machine learning
\citep{li_prioritized_2024,horgan_2018_distributed,schaul_prioritized_2016}.

In nondeterministic environments, trajectory optimization can be implemented by dynamic
programming which has the \MB learner compute $Q$-values as in \MF learning. An
influential perspective on brain modeling combines \MB and \MF learners, sometimes by
arbitrating between them based on measures such as confidence
\citep{daw_uncertainty-based_2005} and sometimes by taking a weighted sum of their
respective $Q$-values \citep{kool_cost-benefit_2017,daw_model-based_2011,Glascher2010}.

While not always explicitly considered, combining \MB and \MF modules is often
motivated by the view that higher-order thinking associated with cortex is more
expensive to use than lower-order subcortical processes. This point is nicely put by
\citep{kool_cost-benefit_2017} who note that in their setup sensibly incorporating a
\MF module ``requires that people assign a cost to \MB control; otherwise, there is
nothing to trade off against its potential.''

Despite the high importance of this topic and the recognized centrality of resource
considerations, memory constraints applied to the \MB learner in this two-system setup
have not been directly investigated. Generally if such a limitation exists, it is
simply set to some value and not explored beyond this (for instance,
\citep{kool_cost-benefit_2017,daw_model-based_2011,daw_uncertainty-based_2005,Glascher2010}
afford the model-based learner the capacity to learn all relevant transitions in the
environment).

An explicit treatment of memory capacity is quite natural and easily motivated, as many
environments animals face will exceed the memory capacity of their higher-order
thinking functions. For instance, imagine an animal takes a sequence of discrete
actions $a$ which nondeterministically move the animal along branching paths to a set
of discrete states $s$. It is valuable to learn the resulting state probability
distribution for each action $P(s'|s,a)$, which amounts to learning a model for the
dynamics of the environment. However, if the number of branches in the environment is
large, the animal will not be able to track all of these probability estimates
explicitly. The animal may give up on using \MB learning altogether and fall back to
\MF; or, as we hypothesize here, the animal may choose a subset of the probabilities
$P(s'|s,a)$ to track.

A third possibility is that the animal uses parametric models to store features that
express combinations of probabilities, or engage in other compression mechanisms such
as the Successor Representation. Recently, the authors of \citep{greenstreet_why_2026}
proposed a model where the \MB learner uses self-supervised learning to learn a
parameterized low-dimensional embedding of $P(s'|s,a)$, which is akin to a
parameterized version of the tabular-RL-oriented memory constraint considered here. The
\MF learner is then trained on this low-dimensional embedding of actions. The
low-dimensional embedding induces a topography in action space, whereby nearby actions
map to nearby embeddings, a property which is seen in recordings from striatum. This
study differs from our approach in at least four key ways: (1) in their model, the
subcortex is fully dependent on the cortex to function, which may not always be the
case in biology \citep{kawai_motor_2015}; (2) they do not treat the dimension of the
embedding space as a parameter and investigate the impact of changing it; (3) they
allow the optimization to determine the low-dimensional embedding, while here we
consider the consequences of different explicitly defined embedding schemes; and (4)
our approach uses tabular reinforcement learning rather than parametric models. Despite
these differences, the compelling framework of \citep{greenstreet_why_2026} would
likely be influential in the development of a parameterized version of our work.

\begin{figure}
  \centering
  \includegraphics[width=\textwidth]{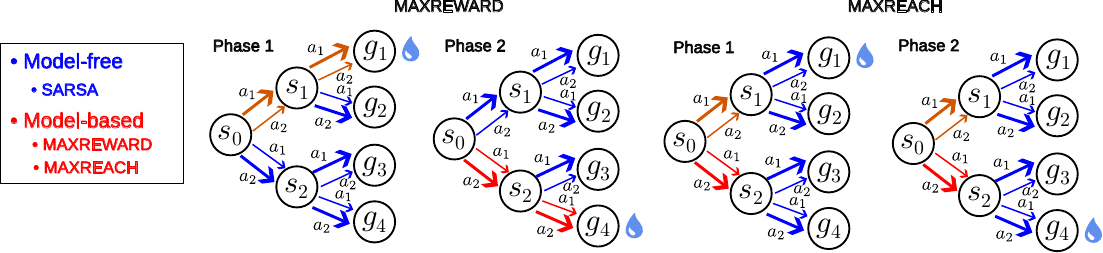}
  \caption{Example learning environment and strategies MAXREWARD and
    MAXREACH, for tree-depth $d=2$ and memory capacity $m=4$. Red arrows denote
    edges tracked by the model-based learner, and blue arrows are edges that use
    only model-free learning. Stroke width of arrows represents the size of the
    probability. The blue water drop represents reward. In this example, for
    MAXREWARD all of the edges that led to reward are eventually tracked,
    while for MAXREACH the edges that are nearest the root node are tracked.}
  \label{fig:mb-strategies}
\end{figure}

\section{Methods}
The learning environment we consider is a tree structure with nondeterministic
transitions from a parent state $s_{\text{parent}}$ to its children, with probability
$P(s_{\text{child}}|s_{\text{parent}},a)$ determined by the action taken $a$. Formally
this tree structure is a directed multi-graph induced by nonzero probabilities, so that
$\edge{a}{s}{s'}$ is an edge in the graph if and only if $P(s'|s,a)>0$, and henceforth
when we use the word ``edge'' we refer to the edges in this graph. The agent always
starts at a root state $s_0$ and eventually arrives at a leaf state, at which point the
episode ends and the agent resets to $s_0$. Reward, if it is given, is only given in
leaf states. A tree environment significantly reduces the complexity of theoretical
analysis and makes the appropriateness of various learning rules more obvious. It also
aligns with many experiments in psychology and neuroscience
\citep{daw_uncertainty-based_2005,daw_model-based_2011,Glascher2010,kool_cost-benefit_2017}.
We consider regular trees here, so that every non-leaf node has two children and at
most two edges connected to each child. The tree is parameterized by the depth $d$,
which is the number of edges that must be passed through to get from the root to a
leaf.

We next need to decide how to define \MF versus \MB learners. It is sometimes assumed
that the \MF learner uses a standard temporal-difference learning algorithm,
corresponding to SARSA(0) \citep{mattar_prioritized_2018}. It is then argued that \MB
control, such as internally simulated trajectories, allows reward information to
propagate backward more quickly toward the starting state. However, it is not clear
that this simple account is the right model for \MF learning, as a relatively simple
``eligibility trace'' such as used in SARSA($\lambda$), or an update sweep that starts
at the rewarded state and works backward, allow reward information to instantaneously
propagate through many states preceding a reward, making the reward-propagating
function of replay somewhat redundant.\footnote{Other hypotheses for simulated
  trajectories via replay are that it implements a learning curriculum which mitigates
  forgetting of important information.}

Our choice is motivated by the following: make the \MF learner relatively powerful,
while considering a simple form of \MB learning. To this aim, we use a SARSA(0)-style
rule for the \MF learner, with a backward-propagating update rule. This means that the
$Q$-function for the last transition $Q(s^{(\text{final}-1)}, a^{(\text{final}-1)})$ in
an episode is first computed, and this updated value is used in the calculation of
$Q(s^{(\text{final}-2)}, a^{(\text{final}-2)})$, and so forth until the first
transition experienced in the episode, $Q(s_{0}, a^{(0)})$. This allows reward
experienced in an episode to be immediately backpropagated in a way similar to
SARSA($\lambda$) while avoiding the addition of an extra parameter $\lambda$. Likewise,
when computing \MB $Q$-values we use dynamic programming with the estimated
probabilities starting from the leaf states and moving backward toward the root. In
summary, both \MF and \MB learners immediately propagate reward information through the
environment back to the root state.

During the first \phaseoneep episodes (Phase 1), the reward pattern is chosen and
fixed. During the last \phaseoneep episodes (Phase 2), the reward shifts to a new
configuration. \fig \ref{fig:mb-strategies} shows an example where only one leaf state
is rewarded in each phase.

Our novel contribution is to constrain the number of environment transition
probabilities $P(s'|s,a)$ that can be tracked and estimated by the \MB learner.
Estimated probabilities will be denoted by $\hat{P}$. The constraint on the number of
edges that can be learned results in different \emph{strategies} for how to choose
these edges. We use $m$ to denote the memory capacity.

Let $E_{s,a}$ denote the set of edges $\edge{a}{s}{\cdot}$ that are being tracked
starting at state $s$ and taking action $a$. To compute $Q^{\text{MB}}(s, a)$, we use
dynamic programming (with the estimated transition probabilities) if tracked edges
$E_{s,a}$ are present and fall back to model-free $Q^{\text{MF}}(s, a)$ otherwise:
\begin{align*}
  V^{\text{MB}}(s)    & = \sum_{a\in A(s)}\pi(a|s)Q^{\text{MB}}(s, a) \\
  Q^{\text{MB}}(s, a) & =
  \begin{cases}
    \sum_{s'\in S} \hat{P}(s'|s,a) (\hat{R}(s') + \gamma V^{\text{MB}}(s')), & E_{s,a} \neq \emptyset \\
    Q^{\text{MF}} (s, a),                                                    & E_{s,a} = \emptyset
  \end{cases}
\end{align*}
where $\emptyset$ is the empty set, $\pi$ is the current policy, $\hat{R}$ is the
estimated reward function deduced from experience, $S$ is the full set of states of the
environment, $A(s)$ are all valid actions in state $s$, and $\gamma$ is the
discount (here set to $1.0$).
If only a strict subset of outgoing edges for $(s,a)$ is tracked, the dynamic-programming
update uses the tracked edges and assigns all untracked probability mass to an implicit
zero-reward terminal absorbing state; it does not fall back to $Q^{\text{MF}}(s,a)$.

Transition probabilities are estimated via
\[
  \hat{P}(s'|s,a) = \frac{\text{count}_{\edge{a}{s}{s'}}(s'|s,a)}{\text{count}_{\edge{a}{s}{s'}}(s,a) + c }.
\]
Here $\text{count}_{\edge{a}{s}{s'}}$ tracks transition statistics for the edge
$\edge{a}{s}{s'}$. For a tracked edge $\edge{a}{s}{s'}$,
$\text{count}_{\edge{a}{s}{s'}}(s'|s,a)$ is the observed number of transitions from $s$
to $s'$ resulting from action $a$ and $\text{count}_{\edge{a}{s}{s'}}(s,a)$ is the
observed number of times action $a$ is taken in state $s$. If the edge
$\edge{a}{s}{s'}$ is no longer tracked, these counts are reset to zero. The constant
$c$ biases probabilities toward zero: for $c>0$ the probability can approach but never
exactly equal $1$. Here we always take $c=1$. Any untracked edges are assumed by the
\MB module to lead to a terminal state that receives no reward. This assumption is
consistent with the model lacking \emph{a priori} knowledge about the structure of the
tree and adopting a pessimistic prior over unmodeled transitions. These estimates are
not true probabilities, as they may not sum to $1$. This non-normalized approach allows
estimates to be locally determined. The policy is taken to be $\epsilon$-greedy with
respect to $Q^{\text{MB}}$, with a value of $\epsilon=0.2$.

We highlight two strategies for allocating memory, called MAXREWARD and MAXREACH. For
MAXREACH, the agent prioritizes edges closest to the root state $s_0$. This choice is
meant to improve the agent's ability to navigate to arbitrary goal states. In a
restricted depth-2 setting with a simplified belief model, we show in the Extended
Results section of the Appendix that knowledge of the four sibling edges at the root is
more valuable than knowledge of four sibling edges at the second level. For MAXREWARD,
the agent prioritizes edges that can be associated with reward, using the closeness to
the root as a tie-breaker. An edge is marked reward-associated if it occurs earlier in
an episode trajectory that subsequently reaches a state with positive $V^{\MF}(s)$ or
receives nonzero reward. Edges not currently marked as reward-associated are removed as
needed to make space for reward-associated edges. The tracked edges corresponding to
these strategies at the end of Phase 1 are indicated by the red arrows in \fig
\ref{fig:mb-strategies}. Note that this is only guaranteed to be accurate
asymptotically as the number of episodes increases to infinity, since the tracked edges
are added based on the agent's experience. At the start of Phase 2
$Q^{\text{MF}}$-values are set to zero and reward associations reset (previously
tracked edges and their counts are retained but their reward association is removed).

See the Extended Methods section of the Appendix for additional details, including
hyperparameter choices (\tabref \ref{tab:hparams}) and full algorithms for MAXREWARD
and MAXREACH. See also \tabref \ref{tab:terms} for a list of symbols used in this
paper.

\begin{figure}
  \centering
  \includegraphics[width=\textwidth]{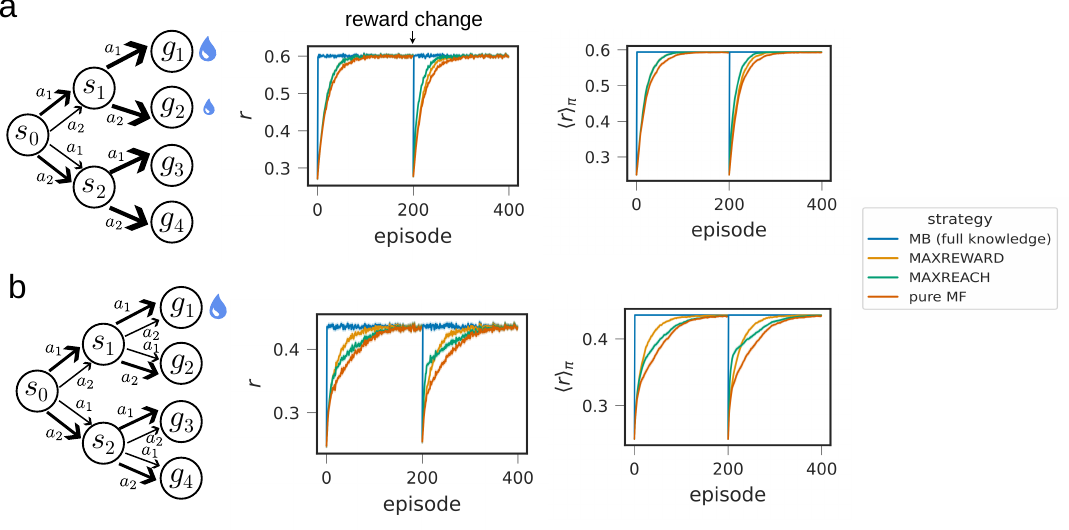}
  \caption{Two illustrative environments comparing the reward $r$ received by
    MAXREWARD and MAXREACH over episodes.
    Legend to the right shows the colors that correspond to different strategies in
    each of the plots. ``MB (full knowledge)'' means dynamic programming using the true
    environment probabilities, and ``pure MF'' means purely model-free. For all
    panels, tree depth is $d=2$ and memory capacity
    is $m=4$. Shaded regions in plots depict standard error of the mean.
    \textbf{(a)} Transition probabilities are $P(s_1|s_0, a_1)=0.7$, $P(s_1|s_0, a_2)=0.3$,
    $P(g_1|s_1, a_1)=1.0$, $P(g_1|s_1, a_2)=0.0$, $P(g_3|s_2, a_1)=1.0$, and
    $P(g_3|s_2, a_2)=0.0$. Other probabilities can be deduced by using that
    probabilities sum to $1$. Reward is $1.0$ for $g_1$ and $0.1$ for $g_2$ in
    Phase 1 and $1.0$ for $g_3$ and $0.1$ for $g_4$ in Phase 2.
    Left: illustration of the environment. Reward (water drop) is shown for Phase 1.
    Center: Reward averaged over 40000 trials, plotted over episodes. In Phase 1
    the lines for MAXREWARD and MAXREACH overlap.
    Right: Reward first averaged over the policy, then averaged over 4000
    trials, plotted over episodes.
    \textbf{(b)} Transition probabilities are $P(s_1|s_0, a_1)=0.7$, $P(s_1|s_0, a_2)=0.3$,
    $P(g_1|s_1, a_1)=0.7$, $P(g_1|s_1, a_2)=0.3$, $P(g_3|s_2, a_1)=0.7$, and
    $P(g_3|s_2, a_2)=0.3$. Other probabilities can be deduced as in \textbf{(a)}.
    Reward is $1.0$ in state $g_1$ in Phase 1 and  $1.0$ in state $g_4$ in Phase 2.
    Left: illustration of the environment. Reward is shown for Phase 1.
    Center: Reward averaged over 40000 trials, plotted over episodes.
    Right: Reward averaged over the policy, then averaged over 4000 trials, plotted over episodes.
  }
  \label{fig:fig2}
\end{figure}

\section{Results}

\subsection{Simple first case}
We compare the performance of different strategies in Fig.~\ref{fig:fig2}. In this
figure, dynamic programming with full knowledge of the environment dynamics, and purely
model-free SARSA, provide upper and lower reference points for performance,
respectively. \fig \ref{fig:fig2}a shows the learning curves for a depth-two tree
($\treedepth=2$) and memory capacity $m=4$ where the second-level transitions are
deterministic; that is, $P(g|s,a)\in \{1,0\}$ where $g$ is a goal state. During Phase 1
a reward value of $1$ and $0.1$ are given if the agent reaches goal state $g_1$ and
$g_2$, respectively, and in Phase 2 this switches to $g_3$ and $g_4$, respectively. In
Phase 1, MAXREWARD will eventually track the edges $\edge{a_1}{s_0}{s_1}$,
$\edge{a_2}{s_0}{s_1}$, $\edge{a_1}{s_1}{g_1}$, and $\edge{a_2}{s_1}{g_2}$, while
MAXREACH will track the edges $\edge{a_1}{s_0}{s_1}$, $\edge{a_2}{s_0}{s_1}$,
$\edge{a_1}{s_0}{s_2}$, and $\edge{a_2}{s_0}{s_2}$ in Phases 1 and 2.

Since the second-level transitions are nearly trivial to model, there is no additional
advantage to using a \MB method for edges $\edge{a_1}{s_1}{g_1}$ and
$\edge{a_2}{s_1}{g_2}$. For this reason, MAXREACH performs just as well as MAXREWARD
during Phase 1; MAXREWARD is essentially ``wasting'' its capacity by modeling the
second-level transitions.

In contrast, MAXREACH does not track the edges $\edge{a_1}{s_1}{g_1}$ and
$\edge{a_2}{s_1}{g_2}$ during Phase 1, but instead estimates the transition
probabilities for $\edge{a_1}{s_0}{s_2}$ and $\edge{a_2}{s_0}{s_2}$. This means that it
has a head-start in the transition from Phase 1 to Phase 2, when the edges
$\edge{a_1}{s_0}{s_2}$ and $\edge{a_2}{s_0}{s_2}$ become relevant for receiving reward.

Instead of directly measuring reward experienced by the agent and averaging this over
trials, we can instead first compute the expected reward given the policy and average
this over trials. In equations,
\[
  \bar{r}(k) = \mathbb{E} \left[ \frac{1}{N}\sum_{n=1}^{N} r_n(k) \right] =  \mathbb{E} \left[ \frac{1}{N} \sum_{n=1}^{N} \mathbb{E}_{\pi_{n,k}, P}[r_n(k)| \mathcal{H}_{n,k}] \right]
\]
where $\overline{r}(k)$ is the average reward at episode $k$, $r_n(k)$ is a random
variable for the reward gained at episode $k$ and trial $n$, and
$\mathbb{E}_{\pi_{n,k}, P}$ is averaging with respect to the policy $\pi_{n,k}$ and
environment transition probabilities $P$ (the policy depends on $n$ and $k$) and
$\mathcal{H}_{n,k}$ is the learning history before episode $k$ in trial $n$. While the
means of both expressions above are the same, the variance of the estimator inside the
expectation is smaller for the latter expression. Intuitively, we have replaced
sampling over the policy choices with the analytical average with respect to the
policy, and removing this sampling reduces the variance. This can be seen directly in
the much cleaner lines of the rightmost panel of \fig \ref{fig:fig2}a compared to the
middle panel, despite using a tenth of the trials. We use this technique going forward
to make the behavior easier to see.

\fig \ref{fig:fig2}a shows a relatively extreme example that strongly favors MAXREACH.
\fig \ref{fig:fig2}b shows a more balanced example where the transition probabilities at
the second level of the depth-2 tree match those of the first level, again for a memory
capacity of $\memcap=4$. Here reward is received only at $g_1$ in Phase 1 and
at $g_4$ in Phase 2. In this case, MAXREWARD tracks the edges
$\edge{a_1}{s_0}{s_1}$, $\edge{a_2}{s_0}{s_1}$, $\edge{a_1}{s_1}{g_1}$, and
$\edge{a_2}{s_1}{g_1}$ in Phase 1, while MAXREACH again tracks the edges
$\edge{a_1}{s_0}{s_1}$, $\edge{a_2}{s_0}{s_1}$, $\edge{a_1}{s_0}{s_2}$, and
$\edge{a_2}{s_0}{s_2}$ in Phases 1 and 2. Here we see that MAXREWARD has the
advantage over MAXREACH during Phase 1, while during Phase 2 MAXREACH initially has an
advantage over MAXREWARD and then is surpassed by MAXREWARD after about 25 episodes.

\subsection{MAXREWARD and MAXREACH across \texorpdfstring{$m$}{m} and \texorpdfstring{$d$}{d}}

\begin{figure}
  \centering
  \includegraphics[width=\textwidth]{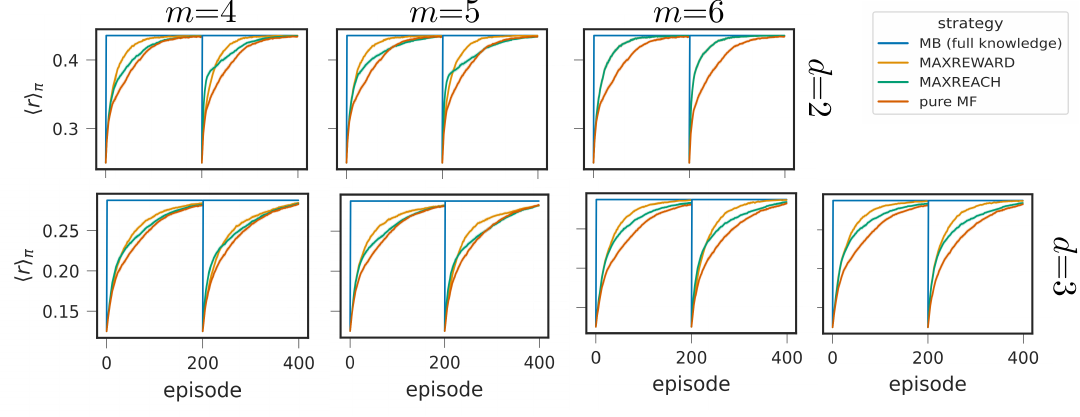}
  \caption{Policy-averaged reward curves $\avg{r}_{\pi}$ for strategies plotted over a range of
    memory capacity $m$ and tree-depth $d$ values. Shaded regions in plots
    depict standard error of the mean.
    Legend on the top right is as in \fig \ref{fig:fig2}. Columns 1 through 4 $m=4$, $5$, $6$, and $8$,
    respectively. Rows 1 and 2:  $d=2$ and $3$, respectively. The plot for $m=8$ and $d=2$
    is identical to that of $m=6$ and $d=2$ and so is omitted.
    For $m=6$ and $d=2$ the MAXREWARD and MAXREACH lines overlap.
  }
  \label{fig:survey}
\end{figure}

A more thorough survey of the performance of the two methods is shown in \fig
\ref{fig:survey}. Here we describe some of the general patterns shown by this figure.
Here when the memory capacity $\memcap$ is sufficiently high, the two strategies of
MAXREWARD and MAXREACH coincide and have the same reward ($d=2$ and $m\in \{6,8\}$).
When memory capacity is limited, MAXREWARD has the advantage in periods where the
reward does not change, such as in Phase 1 and in the later episodes of Phase 2.
However, MAXREACH gains an advantage over MAXREWARD immediately after the reward
location changes. The precise trade-off between the two strategies depends on the
structure of the environment.

\subsection{A framework for modeling experimental data}
We next detail how this theory can be used to elucidate the strategies used by animals
in decision-making tasks.

The main idea is to measure how consistent the actions chosen by a participant in a
decision-making task are with different hypothesized learning strategies. Novelly, we
now allow for the hypothesized learning strategy to use \MB and \MF approaches in
different parts of the environment.

To start, consider a simple experiment to see if participants use a strategy with
elements of MAXREACH. We task the participant with acquiring reward in the tree-like
environments described above. If we use a large enough tree, we can be reasonably
assured that the transition probabilities can't be explicitly learned for every
transition. After learning in Phase 1, the reward then moves to a new location. We can
see the way that the participant responds after finding the reward in the new location
during Phase 2 for the first time. If the participant was learning transition
probabilities close to the root of the tree even if these were not relevant to the
Phase 1 reward, then we can expect them to take the most rewarding action in Phase 2
more quickly. This prediction is both spatial and temporal -- action choice improvement
will be most pronounced immediately after the new reward location is discovered and
nearest the root of the tree. In contrast, if the participant is using a MAXREWARD
style strategy, then learning in Phase 2 would look similar to learning in Phase 1.

We can test this experiment in a fine-grained way, at the level of individual
state-action pairs over individual episodes. Given a history of episode trajectories
(states visited and actions taken), and given a current action $a$ chosen by the
participant in a state $s$, we can measure how consistent this action is with various
hypotheses, such as MAXREWARD and MAXREACH. A similar analysis (without considering
memory-constraints) which aggregates over episodes and focuses on the first-stage
choice can be found in \citep{daw_model-based_2011}.

To test this analysis technique, we generate episode trajectories DATA using a given
strategy STRAT\_DATA. We then train a second strategy STRAT using these episode
trajectories. On episode $k$ and trial $n$ we compute the greedy-optimal actions in
state $s$ predicted by STRAT. If the observed action $a$ seen in DATA is the same as
one of these greedy-optimal actions, then we record a value of $1$. Otherwise, the
value is recorded as a $0$. The average value of this data over trials then yields our
estimate for the \emph{consistency}, which we write $C_{(\text{STRAT}, \text{DATA})}$.
If DATA is generated by a strategy STRAT, then we abuse notation slightly and also
write $C_{(\text{STRAT}, \text{STRAT\_DATA})}$.

\begin{figure}
  \centering
  \includegraphics[width=\textwidth]{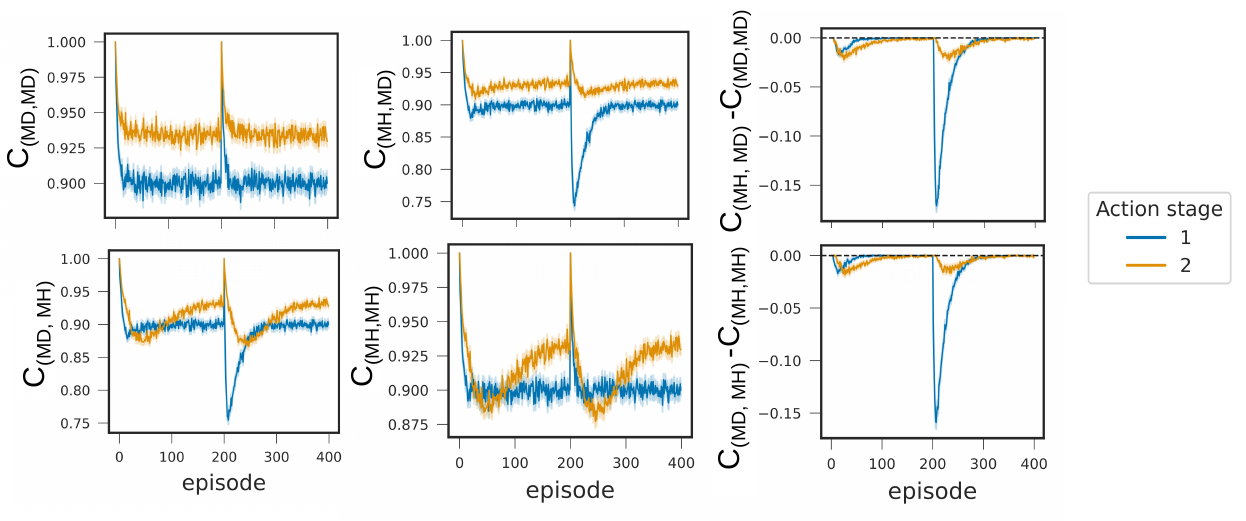}
  \caption{Consistency of MAXREWARD and MAXREACH with themselves and each
    other. Here MD stands for MAXREWARD and MH stands for MAXREACH (first and
    last letter abbreviation),
    and $C_{(\text{STRAT1}, \text{STRAT2})}$ is the consistency of STRAT1 with the data
    produced by STRAT2. The blue line depicts consistency with the
    first action taken in an episode (the first level of the tree) while the
    orange line corresponds with the second action taken (second level of the
    tree). The environment and strategies are as in \fig \ref{fig:fig2}b. For all pairs
    the consistency $C$ is high (roughly between 1.0 and 0.9) across episodes,
    with one exception: the consistency
    of the first action taken after the transition to Phase 2 decreases
    dramatically for $C_{(\text{MD}, \text{MH})}$ and $C_{(\text{MH}, \text{MD})}$,
    and then gradually returns to around 0.9.
    Rightmost column depicts the consistencies $C_{(\text{STRAT1}, \text{STRAT2})}$
    with the baselines $C_{(\text{STRAT2}, \text{STRAT2})}$ subtracted.
    Shaded regions in plots depict standard error of the mean.
  }
  \label{fig:consist}
\end{figure}

\fig \ref{fig:consist} shows the consistency of MAXREWARD and MAXREACH.
The consistencies of MAXREWARD with itself, and MAXREACH with itself, are both
high over episodes, hovering between 1.0 and 0.9 and peaking at the beginning of
Phases 1 and 2 (where the policies start off as uniform). The cross-comparison
of MAXREWARD with MAXREACH and vice versa show clear unique signatures.
The consistency is high across episodes, with one exception: the consistency
of the first-stage action decreases dramatically immediately after the
transition to Phase 2, and then gradually returns to around 0.9. This is a
spatially and temporally precise signature of the cross-consistency of
these two strategies.

In practice DATA would correspond to collected experimental data. The consistency of
various strategies such as MAXREWARD and MAXREACH with DATA can then be measured. This
can then reveal at which episodes, and at what action taken in the episodes, a given
strategy is consistent with the data.

\begin{table}[t]
  \centering
  \caption{Key terms and variables}
  \begin{tabular}{ll}
    \toprule
    \textbf{Symbol}                          & \textbf{Description}                                                                    \\
    \midrule
    \MB                                      & model-based                                                                             \\
    \MF                                      & model-free                                                                              \\
    trial                                    & Independent learning process starting from zero                                         \\
    episode                                  & A sequence of steps taken from root to leaf node ($400$ per trial)                      \\
    Phases 1 and 2                           & Sequences of $200$ episodes with a set reward structure                                 \\
    $s, s'$                                  & Current and next state                                                                  \\
    $a$                                      & Action taken in state $s$                                                               \\
    $P(s'|s,a)$                              & True transition probability                                                             \\
    $\hat{P}(s'|s,a)$                        & Estimated transition probability                                                        \\
    $Q^{\text{MB}}(s,a)$                     & Model-based state-action value                                                          \\
    $Q^{\text{MF}}(s,a)$                     & Model-free state-action value                                                           \\
    $V^{\text{MB}}(s)$                       & Model-based state value                                                                 \\
    $R(s)$                                   & True reward function                                                                    \\
    $\hat{R}(s)$                             & Estimated reward function                                                               \\
    $\pi(a|s)$                               & Policy (state-conditioned action selection probability)                                 \\
    $E_{s,a}$                                & \MB tracked edges for $(s,a)$                                                           \\
    $\edge{a}{s}{s'}$                        & Transition edge in graph ($P(s'|s,a)>0$)                                                \\
    $m$                                      & Memory capacity (number of edges that can be tracked)                                   \\
    $d$                                      & Tree depth                                                                              \\
    $c$                                      & Smoothing constant for $\hat{P}$ estimate                                               \\
    $\text{count}_{\edge{a}{s}{s'}}(s'|s,a)$ & Number of transitions tracked for edge $\edge{a}{s}{s'}$ from $s$ to $s'$ by action $a$ \\
    $\text{count}_{\edge{a}{s}{s'}}(s,a)$    & Total times action $a$ taken in $s$ tracked for edge $\edge{a}{s}{s'}$                  \\
    $C_{(\text{STRAT}, \text{DATA})}$        & Consistency of strategy STRAT with data DATA                                            \\
    \bottomrule
  \end{tabular}
  \label{tab:terms}
\end{table}

\section{Discussion}
Here we introduced a model using memory-capacity-limited model-based (\MB) learning
working in concert with a model-free (\MF) learner. This naturally led to the concept
of different strategies for how to allot the limited memory capacity. We considered two
broad and antipodal strategies, MAXREWARD and MAXREACH, which prioritize exploiting the
current reward and maximizing arbitrary goal reaching, respectively. Note that this is
different than the usual exploration-exploitation tradeoff explored in reinforcement
learning: here we are focused on balancing these factors at the level of \MB learning
structure, rather than at the level of action selection as is usually done.

We found that MAXREACH can have an advantage in the time period after the reward
location changes. If this nonstationarity is dominant in the natural setting of
animals, then it may be that they use a strategy that leans toward MAXREACH. This has
several intriguing implications.

First, MAXREACH corresponds to the \MB learner guiding its model-building not based on
reward, but on more general considerations of the environment. With such a model,
reward signals can quickly be used to construct a strategy for exploiting these
signals; there is no need for the repeated reward experience needed by the model-free
system. This means that the model-free and \MB learners receive and use reward
information in different ways. If animals use such a strategy, it may explain why
cortex (\MB) receives sparser dopamine projections (especially relative to other
neurotransmitters such as norepinephrine) when compared to striatum (\MF)
\citep{lopez_innervation_2024, tritsch_dopaminergic_2012}, and may guide further
hypotheses for exploring the nature of these differences. Furthermore, it may help
explain why cortex evolved to be physically distinct and spatially separate from the
basal ganglia; this helps ensure that the reward signals strongly projecting to
striatum are not picked up by the cortex.

Indeed, humans have been observed to learn transition probabilities in the absence of
explicit reward signals \citep{Glascher2010}, though reward is also seen to influence
\MB control \citep{kool_cost-benefit_2017}, and dopamine projections to cortex have
been observed to be a necessary component for learning
\cite{molina-luna_dopamine_2009}. This aligns reasonably well with our model, where the
\MB module needs reward signals to compute $Q$ values but does not rely on reward to
continue learning useful structural information from the environment, while the \MF
module requires reward signals be repeatedly provided to learn.

Here we propose to investigate if transition probabilities that are not associated with
reward are learned even when the number of transition probabilities that are associated
with reward exceeds a person's ability to remember. Experimental evidence indicates
that cortex responds to violations of expected stimuli that are not reward-associated
\citep{courchesne_stimulus_1975,fabiani_changes_1995}, unlike dopamine neurons that are
largely concerned with reward \citep{holroyd_neural_2002}, though recent work has
challenged this understanding of dopamine (for a review see
\citep{kahnt_curious_2025}). A more nuanced picture would incorporate where differently
tuned dopamine neurons project to; as our model suggests, it may be that reward-tuned
dopamine neurons project primarily to subcortical structures like striatum, while
sensory-prediction tuned dopamine projects primarily to cortical structures.

Another intriguing possibility is that the cortex transiently allots memory resources
to learning a new task by seeking to exploit reward, but that these memory slots are
pruned as subcortical learning catches up. This would correspond to adding a hot cache
to MAXREACH to create a new strategy that balances MAXREACH and MAXREWARD. Indeed,
spines in cortex have been observed that closely match this hypothesis; they are
transiently formed in response to a new rewarded task, and subsequently pruned
\citep{xu_rapid_2009}.

Finally, we proposed a method to test these hypotheses concretely in experiments,
defining an analysis of consistency between data and hypothesized learning strategies.
We show how this can be applied in a temporally and spatially precise way, teasing out
specific signatures of the learning strategies that may be used by animals.

Our model has several limitations and opportunities for future development. First of
all, our setting is a toy environment with simple algorithms. Since we are focused on
modeling the brain and in laying a framework for a theory that can be easily
understood, we believe that this setting is a reasonable first start. However,
especially when applying our findings to machine learning, it is important to see if
our findings scale to more complex environments and larger and more sophisticated
models. One important direction is that of parametric models in contrast to the tabular
methods we explored here.

Another limitation is that we did not fully explore the ways that the \MF and \MB
learning signals can be knit together. Indeed, our \MB learner can underestimate reward
signals, since untracked edges are assigned a probability of $0$. We believe that it
would be a fruitful research direction to consider cleverer ways of combining these
signals to avoid some of these shortcomings.

We also did not explicitly address the matter of \MB control being handed over to \MF
control after the task has been mastered. In our setting there is no penalty to using
the tracked edges to compute the \MB $Q$ values, so there is not an impetus for phasing
out this computation. Adding such a computation cost would allow our model to extend to
capturing this phenomenon (observed in experiments \citep{kawai_motor_2015}) wherein
the model relies fully on the \MF learner asymptotically (at least until there is a
transition to a novel environment/reward structure).

We also did not explicitly consider the case where animals return to a previously
familiar environment/reward structure. In our model, the \MF and \MB learners largely
erase information about previous reward location. A more realistic model would
incorporate the more persistent storage of solutions derived for previously encountered
environments. In general there are many other reasonable strategies to consider beyond
MAXREACH and MAXREWARD. We focused on these two as they are two extremes of a spectrum,
but an intermediate approach (such as the idea of a hot cache as discussed above) would
likely be more faithful to animal brains and practically useful for machine learning.

Cortical higher-order function has many facets beyond what is explored here. Here we
assume the \MB learner is used to learn the dynamics underlying the environment.
Trajectory simulation, imitation learning, task decomposition (inferring a different
set of state-action values), and more are likewise interesting types of \MB learning
that could be enriched with this memory-capacity-limitation framework.

\clearpage

\section*{Code availability}
Code is being prepared for release at a later date.

\section*{Acknowledgements}
M.F. was supported by the Special Postdoctoral Researchers Program at
RIKEN.
T.T. was supported by RIKEN Center for Brain Science, RIKEN TRIP initiative
(RIKEN Quantum), JST CREST Grant Number JPMJCR23N2, and JSPS KAKENHI Grant
Number JP25K24466.

\appendix
\section{Appendix}
\subsection{Extended Methods}
\begin{table}[t]
  \centering
  \caption{Hyperparameters used in experiments.}
  \begin{tabular}{lll}
    \toprule
    \textbf{Symbol} & \textbf{Name}                 & \textbf{Value / Description}           \\
    \midrule
    $\epsilon$      & Exploration rate              & $0.2$ (for $\epsilon$-greedy policies) \\
    $\alpha$        & Step size for \MF learner     & $0.1$                                  \\
    $\gamma$        & Discount rate                 & $1.0$                                  \\
    $c$             & Transition smoothing constant & $1$                                    \\
    $m$             & Memory capacity               & Varies                                 \\
    $d$             & Tree depth                    & Varies                                 \\
                    & Phase 1                       & Episodes $1$ through $200$             \\
                    & Phase 2                       & Episodes $201$ through $400$           \\
                    & Compute cores for simulations & 20                                     \\
                    & Memory used for simulations   & $\sim$60 GB                            \\
                    & Compute time required         & $\sim$1 hour                           \\
    \bottomrule
  \end{tabular}
  \label{tab:hparams}
\end{table}

Our learning environment is a probabilistic sequential Markov decision process (MDP)
with a tree topology (a directed multi-graph). The learning agent always starts an
episode in the root node $s_0$. Leaf nodes in the tree are terminal -- when the agent
reaches a leaf-node state, the episode ends and the next episode starts. In each
non-terminal state $s$, the agent can choose one of two actions $a_1$ or $a_2$. After
choosing an action the agent transitions to a new state $s'$ according to a probability
$P(s'|s,a)$. This probability can be nonzero only if $s'$ is a child of $s$ in the
tree. We consider the case where every non-leaf node has exactly two children. Reward
may only be received upon reaching a leaf node. The amount of reward received in each
leaf node is split into two phases, Phases 1 and Phase 2, each lasting $200$ episodes.
Phase 2 immediately follows Phase 1.

Our agent consists of two modules, one model-based (\MB) and the other model-free
(\MF). Each module computes $Q$-values, denoted $Q^{\MB}$ and $Q^{\MF}$, respectively.
$Q$ values assign a value to state-action pairs $Q(s, a)\in \mathbb{R}$, and are a
standard construct in reinforcement learning.

Our \MF module computes reward according to a SARSA(0) algorithm, with one twist: $Q$
values are computed starting at the leaf nodes and moving toward the root, and updated
$Q$ values are used in the computation of subsequent $Q$ values (backward sweep). This
means that experiencing reward in an episode assigns a nonzero increase in the $Q$
values for all states and actions that occurred during that episode. SARSA(0) has two
hyperparameters: $\gamma$ and $\alpha$. The scalar $\gamma$ is the usual
reinforcement-learning discount factor, and we take it to be $\gamma = 1.0$. The scalar
$\alpha$ is the step-size, which we take to be $\alpha = 0.1$ (with the main update
step highlighted). Pseudo-code for updating $Q$ based on an episode is given in \algr
\ref{alg:backward-sarsa}. Note that reward, as it propagates from the leaf to the root,
is attenuated by factors of $\alpha$ up to $\alpha^{d}$ (a different scaling than
forward-sweep SARSA($\lambda$)). For deeper trees or more complex environments this may
be relevant, but for the $d=2,3$ trees we consider here this attenuation should not
hamper the \MF learner's ability to learn the reward structure.

\begin{algorithm}[t]
  \caption{SARSA: Backward SARSA-style update over one trajectory}
  \label{alg:backward-sarsa}
  \begin{algorithmic}[1]
    \Require Episode trajectory $\tau = \{(s^{(0)},a^{(0)},r^{(1)}), (s^{(1)},a^{(1)},r^{(2)}), \ldots,
      (s^{(T-1)},a^{(T-1)},r^{(T)}), s^{(T)}\}$, $Q$, learning rate $\alpha$, discount factor $\gamma$
    \Statex \textbf{Notation:} $s^{(k)}$ is the state at time step $k$, $a^{(k)}$
    is the action taken in $s^{(k)}$, and $r^{(k)}$ is the reward observed after
    transitioning into $s^{(k)}$

    \State Set $Q(s^{(T)},a) \gets 0$ for all actions $a$

    \For{$t = T-1, T-2, \ldots, 0$}
    \State $s \gets s^{(t)}$
    \State $a \gets a^{(t)}$
    \State $r \gets r^{(t+1)}$
    \State $s' \gets s^{(t+1)}$

    \If{$s'$ is terminal}
    \State $y \gets r$
    \Else
    \State $a' \gets a^{(t+1)}$
    \State $y \gets r + \gamma Q(s',a')$
    \EndIf

    \State $Q(s,a) \gets Q(s,a) + \alpha \bigl(y - Q(s,a)\bigr)$ \qquad (SARSA update)
    \EndFor
    \State \Return $Q$
  \end{algorithmic}
\end{algorithm}

The \MB learner is based on dynamic programming using the learned transition
probabilities in the environment, with the added constraint of limited memory capacity.
This means that only a limited number, $m$, of transition probabilities can be tracked
at any given time. We give the precise rule for how these are determined when
describing the MAXREWARD and MAXREACH strategies below. Note that our implementation of
dynamic programming uses policy-evaluation (the value function $V(s)$ is computed by
averaging the $Q$ values over the policy rather than by taking $V(s)=\max_{a}Q(s,a)$).

The model-based learner uses dynamic programming which also follows a backward sweep
from leaf nodes to root node. An important caveat is that this model can only use
transition probabilities that it is tracking in its memory. Let $E_{s,a}$ denote the
set of edges $\edge{a}{s}{\cdot}$ that are being tracked starting at state $s$ and
taking action $a$. We use the dynamic programming update if tracked edges $E_{s,a}$ are
present and fallback to model-free $Q^{\text{MF}}(s, a)$ otherwise. The precise rule is
seen in \algr \ref{alg:backward-dp}. Note that this can underestimate reward signals
since missing edges are treated as having probability $0$.

\begin{algorithm}
  \caption{DP: Dynamic programming on tracked edges, backward sweep (tree MDP), and \MF fallback}
  \label{alg:backward-dp}
  \begin{algorithmic}[1]
    \Require Tree MDP $(S, A, P, R, \gamma)$, tracked edges $E_{s,a}$ for all $s$ and
    $a$, $Q^{\text{MF}}$, $\pi$
    \State \textbf{Notation:} Note that $P$ may be an estimate of the true transition probabilities rather than the truth. $R$ may also be an estimate of the true reward function.
    \State Sort $S$ from leaf-most to root-most states, with ties broken arbitrarily
    \State Initialize $V$
    \State Initialize $Q$
    \ForAll{$s \in S$}
    \State $V(s) \gets 0$
    \EndFor
    \For{$s \in S$} \Comment{Iterate with respect to $S$'s sorting (leaf nodes first)}
    \For{each $a \in A(s)$}
    \If{$E_{s,a} = \emptyset$} \Comment{No tracked edges for $(s,a)$}
    \State $Q(s,a) \gets Q^{\text{MF}}(s,a)$
    \Else \Comment{Dynamic programming policy evaluation}
    \State $Q(s,a) \gets \sum_{s'\in S} P(s' \mid s,a)\left[ R(s') + \gamma V(s') \right]$
    \EndIf
    \EndFor
    \State $V(s) \gets \sum_{a\in A(s)} \pi(a \mid s)\, Q(s,a)$ \Comment{Zero if $A(s)$ is empty.}
    \EndFor
    \State \Return $Q$
  \end{algorithmic}
\end{algorithm}

During the first \phaseoneep episodes (Phase 1), the reward pattern is chosen and fixed
in a way that is specified on a per-experiment basis. During the last \phaseoneep
episodes (Phase 2), the reward shifts to a new configuration.

We next formally define our MAXREWARD and MAXREACH strategies. For this we need to
define the \MF value function $V^{\text{MF}}(s) = \sum_{a} \pi(a|s) Q^{\text{MF}}(s,
  a)$. The simple way to understand MAXREWARD is that it favors edges that are related to
reward. We define an edge as reward-associated if, after traversing it in an episode
trajectory, the trajectory reaches a state with $V^{\text{MF}}(s)>0$ or receives
nonzero reward. MAXREACH instead favors edges that maximize the ability to reach
arbitrary goals (see A.2 Extended Results for more information). This roughly
corresponds to edges with the most descendant leaf nodes (equivalent in our case to the
edges closest to the root). The other source of complexity comes from deciding
tie-breaks. The algorithm is written in detail in \algr \ref{alg:strats}.

\begin{algorithm}[t]
  \caption{EDGESELECT: MAXREWARD and MAXREACH edge selection}
  \label{alg:strats}
  \begin{algorithmic}[1]
    \Require Episode trajectory $\tau = \{(s^{(0)},a^{(0)},r^{(1)}), (s^{(1)},a^{(1)},r^{(2)}), \ldots,
      (s^{(T-1)},a^{(T-1)},r^{(T)}), s^{(T)}\}$, $Q^{\text{MF}}$, $E$, $L$, $V^{\text{MF}}$,
    LEAFCOUNT, STRATEGY, $m$, count
    \Statex \textbf{Notation:} $E$ is the set of currently tracked edges
    $E=\bigcup_{s,a} E_{s,a}$. $L$ is the set of all transitions currently associated with reward.
    The update rule for $L$ is given below.
    LEAFCOUNT is a function from states $s$ to the integers that returns the number of
    descendant leaf states of $s$ in the tree. STRATEGY is a string specifying the strategy to use:
    STRATEGY$\in\{$"MAXREWARD", "MAXREACH"$\}$. $m$ is the chosen memory capacity.
    \For{$t = T, T-1,\ldots, 1$}
    \If{$V^{\text{MF}}(s^{(t)}) > 0$ OR $r^{(t)} > 0$}
    \For{$t' = t, t-1, t-2,\ldots, 1$}
    \State $L \gets L \cup \{e_{a^{(t'-1)}}(s^{(t'-1)},s^{(t')}) \}$ \Comment{Add reward-associated edges in $\tau$ to $L$}
    \EndFor
    \EndIf
    \EndFor

    \State $E^{\text{prev}} \gets E$
    \For{$t = 0, 1,\ldots T-1$}
    \State $s \gets s^{(t)}$
    \State $s' \gets s^{(t+1)}$
    \State $a \gets a^{(t)}$
    \State $E \gets E \cup \{\edge{a}{s}{s'}\}$ \Comment{Add edges in $\tau$ to $E$}
    \ForAll{$\tilde{s}$ such that $e_a(s,\tilde{s}) \in E$}
    \State $\text{count}_{e_a(s,\tilde{s})}(s,a) \gets
      \text{count}_{e_a(s,\tilde{s})}(s,a) + 1$
    \EndFor
    \State $\text{count}_{\edge{a}{s}{s'}}(s'|s,a) \gets
      \text{count}_{\edge{a}{s}{s'}}(s'|s,a) + 1$
    \EndFor

    \While{|E| > $m$} \Comment{If memory capacity exceeded, remove edges according to STRATEGY}
    \State Scores $\gets \{\}$
    \ForAll{$e_{a}(s, s')\in E$}
    \If{$e_{a}(s, s') \in L$}
    \State $w_1 \gets 1$
    \Else
    \State $w_1 \gets 0$
    \EndIf
    \State $w_2 \gets \text{LEAFCOUNT}($s$)$
    \State $w_3 \gets V^{\text{MF}}(s')$
    \If{$e_{a}(s, s') \in E^{\text{prev}}$}
    \State $w_4 \gets 1$ \Comment{Favor already tracked edges, other factors equal}
    \Else
    \State $w_4 \gets 0$
    \EndIf
    \State Draw $w_5 \sim \mathrm{Uniform}(0,1)$ \Comment{Breaks any remaining ties}
    \If{STRATEGY == "MAXREWARD"}
    \State Scores$[e_{a}(s, s')] \gets (w_1, w_2, w_3, w_4, w_5)$
    \ElsIf{STRATEGY == "MAXREACH"}
    \State Scores$[e_{a}(s, s')] \gets (w_2, w_1, w_3, w_4, w_5)$
    \Else
    \State \textbf{error} \Comment{invalid STRATEGY}
    \EndIf
    \EndFor
    \State $e^{\star}_{a}(s,s') \gets \arg\min\limits_{e \in E} \text{Scores}[e]$
    \Comment{lexicographical order minimum-scoring edge}
    \State Let $e^\star = e_{a^\star}(s^\star, {s'}^\star)$
    \State Remove $e^\star$ from $E$
    \State $\text{count}_{e^\star}({s'}^\star|s^\star,a^\star) \gets 0$
    \State $\text{count}_{e^\star}(s^\star,a^\star) \gets 0$
    \EndWhile

    \State \Return $E$, $L$, count
  \end{algorithmic}
\end{algorithm}

Our complete learning process now combines \algrs \ref{alg:backward-sarsa},
\ref{alg:backward-dp}, and \ref{alg:strats}, which is written out in \algr
\ref{alg:full}.

\begin{algorithm}[t]
  \caption{MAXREWARD and MAXREACH full learning process (one trial)}
  \label{alg:full}
  \begin{algorithmic}[1]
    \Require Tree MDP $(S, A, P, R, \gamma)$, SARSA (\algr
    \ref{alg:backward-sarsa}), DP (\algr \ref{alg:backward-dp}), EDGESELECT
    (\algr \ref{alg:strats}), STRATEGY, LEAFCOUNT, $m$, $\alpha$, $c$, $\epsilon$
    \State Initialize count (all values $\text{count}_{\edge{a}{s}{s'}}(s'|s,a)=0$ and $\text{count}_{\edge{a}{s}{s'}}(s,a)=0$ for valid edges $\edge{a}{s}{s'}$), $Q^{\text{MF}}(s, a)\gets 0$, $\pi(a | s)\gets 1/|A(s)|$ for all non-leaf $s\in S$ and $a\in A(s)$, $\hat{R}(s)\gets 0$ for all $s$
    \State Initialize $E \gets \emptyset$, $L \gets \emptyset$
    \State return\_stats $\gets$ \{\}
    \State return\_stats[``trajectories''] $\gets$ [\ ],
    \State return\_stats[``Q\_MB\_values''] $\gets$ [\ ]
    \For{$n = 0, 1,\ldots 199$} \Comment{Phase 1}
    \State $s \gets s_0$
    \State $\tau \gets [\ ]$
    \While{$s$ is not a leaf node}
    \State Draw $a \sim \pi(\cdot | s)$ \Comment{Execute an episode trajectory}
    \State Take action $a$, observe reward $r$ and next state $s' \sim P(\cdot \mid s,a)$
    \State $\tau.\mathrm{append}((s, a, r))$
    \State $\hat{R}(s') \gets r$
    \State $s \gets s'$
    \EndWhile
    \State $\tau.\mathrm{append}(s)$
    \State $Q^{\text{MF}} \gets \text{SARSA}(\tau, Q^{\text{MF}}, \alpha, \gamma)$
    \ForAll{$s\in S$}
    \State $V^{\text{MF}}(s) \gets \sum_{a\in A(s)} \pi(a|s) Q^{\text{MF}}(s, a)$
    \EndFor
    \State $E, L, \text{count} \gets \text{EDGESELECT}(\tau, Q^{\text{MF}}, E, L, V^{\text{MF}},
      \text{LEAFCOUNT}, \text{STRATEGY}, m, \text{count})$
    \ForAll{$s, s'\in S, a \in A$}
    \If{$\edge{a}{s}{s'} \in E$}
    \State $\hat{P}(s'|s,a) \gets \frac{\text{count}_{\edge{a}{s}{s'}}(s'|s,a)}{\text{count}_{\edge{a}{s}{s'}}(s,a) + c}$
    \Else
    \State $\hat{P}(s'|s,a) \gets 0$
    \EndIf
    \EndFor
    \State $Q^{\text{MB}} \gets \text{DP}(S, A, \hat{P}, \hat{R}, \gamma, E, Q^{\text{MF}}, \pi)$
    \ForAll{$s\in S$ such that $A(s)\neq \emptyset$} \Comment{Now make $\epsilon$-greedy policy}
    \State $A_s \gets A(s)$
    \State $q_{\max} \gets \max\limits_{a\in A_s} Q^{\text{MB}}(s,a)$
    \State $M_s \gets \{a\in A_s : Q^{\text{MB}}(s,a) \approx q_{\max}\}$ \Comment{Tied maximizers}
    \State $K_s \gets |M_s|$, \quad $N_s \gets |A_s|$
    \ForAll{$a\in A_s$}
    \If{$a\in M_s$}
    \State $\pi(a|s) \gets (1-\epsilon)\frac{1}{K_s} + \epsilon\frac{1}{N_s}$
    \Else
    \State $\pi(a|s) \gets \epsilon\frac{1}{N_s}$
    \EndIf
    \EndFor
    \EndFor
    \State return\_stats[``trajectories''].append($\tau$)
    \State return\_stats[``Q\_MB\_values''].append($Q^{\text{MB}}$)
    \EndFor \Comment{End Phase 1}

    \State Update $R$ to the Phase 2 reward function.
    \State $Q^{\text{MF}}(s, a)\gets 0$, $\pi(a | s)\gets 1/|A(s)|$ for all non-leaf $s\in S$ and $a\in A(s)$
    \State $L \gets \emptyset$, $\hat{R}(s)\gets 0$ for all states $s$. \Comment{$E$ and count are retained from Phase 1.}
    \For{$n = 200, 201,\ldots 399$} \Comment{Phase 2}
    \State Repeat steps for Phase 1
    \EndFor \Comment{End Phase 2}
    \State Return return\_stats

  \end{algorithmic}
\end{algorithm}

Simulations were run using 20 CPU cores and around 60GB of RAM. The number of trials
can be reduced if less RAM is available. With these resources the simulations should
take around an hour.

\clearpage

\subsection{Extended Results}
\paragraph{Example: tracking rootward branches facilitates reaching arbitrary leaf nodes.}
Here we argue in a specific limited setting that the MAXREACH strategy facilitates
reaching arbitrary leaf nodes. This analysis uses a simplified goal-conditioned setting
in which untracked transitions are treated as uniform under the belief model. This
differs from the simulation model above, where untracked mass is treated as leading to
a no-reward terminal state and empty tracked-edge sets fall back to model-free values.
We also only compare in the case where all sibling edges are tracked together in
bunches of four, not in the general case where edges can be tracked arbitrarily as
considered in the simulation model.

Consider a depth-2 binary tree Markov Decision Process (MDP) with root state \(s_0\),
two intermediate states \(s_L,s_R\), and four terminal leaf-node states
\[
  \mathcal G=\{LL,LR,RL,RR\}.
\]
At every non-terminal state the agent has two available actions,
\[
  \mathcal A=\{a_1,a_2\}.
\]
The true transition probabilities are assumed to be the same at every branch:
\[
  P(\text{left child}\mid s,a_i)=p_i,
  \qquad
  P(\text{right child}\mid s,a_i)=1-p_i,
\]
for \(i\in\{1,2\}\). Define
\[
  \delta := |p_1-p_2|.
\]

We consider the goal-conditioned control problem in which the agent is given a terminal
goal \(G\in\mathcal G\) before acting. The agent then chooses actions according to its
internal belief model to attempt to reach the goal. Performance (ability to reach the
goal) is evaluated under the true transition dynamics.

For a belief model \(\hat{P}\), let \(\pi_{\hat{P}}(\cdot\mid G)\) denote a
goal-conditioned policy that maximizes the agent's believed probability of reaching
goal \(G\). This policy $\pi_{\hat{P}}$ may not be uniquely determined; to make it
unique, we take the policy that chooses uniformly at random among actions that are
optimal under the belief model (the maximum-entropy policy among conforming policies).
The true goal-conditioned success probability is
\[
  U_{\hat{P}}(G)
  :=
  \mathbb{P}\bigl(S^{(2)}=G \mid \pi_{\hat{P}}(\cdot\mid G), P\bigr),
\]
where $P$ denotes the true transition dynamics and $S^{(2)}$ is a random variable for
the final state the agent reaches. We evaluate the expected goal-conditioned success
probability under a uniform distribution over terminal goals:
\[
  U_{\hat{P}}
  :=
  \frac{1}{4}\sum_{G\in\mathcal G} U_{\hat{P}}(G).
\]

\paragraph{Useful one-step quantity.}

At a known binary branch, if the agent wants to move left, it chooses the action with
larger left-transition probability. If it wants to move right, it chooses the action
with smaller left-transition probability. Hence the best achievable probability of
moving in the desired direction is
\[
  \max\{p_1,p_2\}
\]
for a left target, and
\[
  \max\{1-p_1,1-p_2\}
  =
  1-\min\{p_1,p_2\}
\]
for a right target.

Averaging over left and right targets gives
\[
  \beta
  :=
  \frac{1}{2}\max\{p_1,p_2\}
  +
  \frac{1}{2}\bigl(1-\min\{p_1,p_2\}\bigr).
\]
Since
\[
  \max\{p_1,p_2\}-\min\{p_1,p_2\}=|p_1-p_2|=\delta,
\]
we obtain
\[
  \beta
  =
  \frac{1+\delta}{2}.
\]

\paragraph{Case 1: first-level transitions known, second-level transitions unknown.}

Suppose the agent knows the root-level transition probabilities, but assumes that all
second-level transitions are uniform. That is, at the root the agent knows
\[
  P(L\mid s_0,a_i)=p_i,
  \qquad
  P(R\mid s_0,a_i)=1-p_i,
\]
but at each intermediate state $s$ it believes
\[
  \hat{P}(\text{left child}\mid s,a_i)
  =
  \hat{P}(\text{right child}\mid s,a_i)
  =
  \frac12.
\]
This satisfies a memory capacity of $\memcap=4$.

Because the agent knows the first-level dynamics, it can choose the root action so as
to maximize the probability of reaching the correct half of the tree. Averaged over
goals, this succeeds with probability \(\beta\). Once in the correct intermediate
state, the agent has no useful second-level information, so according to our maximum
entropy assumption the policy chooses actions uniformly. The probability of reaching
the correct terminal child from the middle level is thus \(1/2\). Therefore
\[
  U_{\text{first-known}}
  =
  \beta\cdot \frac12.
\]
Using \(\beta=(1+\delta)/2\), we get
\[
  U_{\text{first-known}}
  =
  \frac{1+\delta}{4}.
\]

\paragraph{Case 2: first-level transitions unknown, one second-level branch known.}

Now suppose the agent believes that the first-level transition is uniform:
\[
  \hat{P}(L\mid s_0,a_i)
  =
  \hat{P}(R\mid s_0,a_i)
  =
  \frac12,
\]
for both actions \(a_i\). However, the agent knows the transition probabilities in one
second-level branch. Suppose the known branch is the left intermediate state \(s_L\).
Thus, for \(s_L\), the agent knows
\[
  P(LL\mid s_L,a_i)=p_i,
  \qquad
  P(LR\mid s_L,a_i)=1-p_i,
\]
while for \(s_R\) it assumes uniform transitions:
\[
  \hat{P}(RL\mid s_R,a_i)
  =
  \hat{P}(RR\mid s_R,a_i)
  =
  \frac12.
\]
As in Case 1, this satisfies a memory capacity of $\memcap=4$.

Because the agent believes the root transition is uniform, its root action does not
affect the believed probability of reaching \(s_L\) or \(s_R\). Under our
maximum-entropy assumption the policy will choose among these actions with uniform
probability. Under the true dynamics, the probability of reaching the left branch is
therefore
\[
  \eta
  :=
  \frac12 p_1+\frac12 p_2
  =
  \frac{p_1+p_2}{2}.
\]
The probability of reaching the right branch is \(1-\eta\).

For goals in the known left branch, which make up half of the possible terminal goals,
the agent reaches the left branch with probability \(\eta\), and then can use its
second-stage knowledge to reach the correct terminal child with average probability
\(\beta\). Hence the contribution from left-branch goals is
\[
  \frac12 \eta \beta.
\]

For goals in the unknown right branch, which also make up half of the possible terminal
goals, the agent reaches the right branch with probability \(1-\eta\), but has no
useful second-stage information. Under the maximum-entropy assumption, it chooses
uniformly at random and reaches the correct terminal child with probability \(1/2\).
The contribution from right-branch goals is
\[
  \frac12 (1-\eta)\frac12
  =
  \frac{1-\eta}{4}.
\]

Therefore
\[
  U_{\text{left-second-known}}
  =
  \frac{\eta\beta}{2}
  +
  \frac{1-\eta}{4}.
\]
Substituting \(\beta=(1+\delta)/2\), we obtain
\[
  U_{\text{left-second-known}}
  =
  \frac{\eta(1+\delta)}{4}
  +
  \frac{1-\eta}{4}
  =
  \frac{1+\eta \delta}{4}.
\]

If instead the known second-level branch is the right branch \(s_R\), then the same
argument gives
\[
  U_{\text{right-second-known}}
  =
  \frac{(1-\eta)\beta}{2}
  +
  \frac{\eta}{4}
  =
  \frac{1+(1-\eta)\delta}{4}.
\]

\paragraph{Comparison.}

The advantage of knowing the first-level transitions rather than the left second-level
branch is
\[
  U_{\text{first-known}}
  -
  U_{\text{left-second-known}}
  =
  \frac{1+\delta}{4}
  -
  \frac{1+\eta \delta}{4}
  =
  \frac{(1-\eta)\delta}{4}.
\]

Since \(\delta\ge 0\) and \(1-\eta\ge 0\), we have
\[
  U_{\text{first-known}}
  \ge
  U_{\text{left-second-known}}.
\]
The inequality is strict whenever
\[
  p_1\neq p_2
\]
and
\[
  \eta < 1.
\]

The advantage of knowing the first-level transitions rather than the right second-level
branch is
\[
  U_{\text{first-known}}
  -
  U_{\text{right-second-known}}
  =
  \frac{1+\delta}{4} - \frac{1+(1-\eta)\delta}{4}
  =
  \frac{\eta \delta}{4}.
\]
Again,
\[
  U_{\text{first-known}}
  \ge
  U_{\text{right-second-known}}.
\]

Thus, under expected goal-conditioned success probability, knowing the first-level
transition dynamics weakly dominates knowing the transition dynamics for only one
second-level branch.

\paragraph{Conclusion.}

In a depth-2 binary tree MDP with regular transition dynamics, transition knowledge at
the first level is more valuable for goal-conditioned control than transition knowledge
restricted to one second-level branch. This is because first-level knowledge helps
route the agent toward the correct half of the tree for every possible terminal goal,
whereas knowledge of a single second-level branch only helps for goals in that branch
and only conditional on reaching it.

\end{document}